\documentstyle[graphicx,twocolumn,aps,floats]{revtex}
\begin{document}

\title{Evidence for superfluid B-phase of $^3$He in aerogel}
\draft
\author{H.  Alles\cite{newadd}, J.  J.  Kaplinsky, P.  S.  Wootton, J.
D.  Reppy\cite{newadd1}, J.  H.  Naish and J.R.  Hook}
\address{Schuster Laboratory, University of Manchester, Manchester,
M13 9PL, U.K.}
\date{\today}
\maketitle

\begin{abstract}

We have made simultaneous torsional oscillator and transverse cw NMR
(at $\sim165\,{\rm kHz}$) studies of the superfluid phase of $^3$He in
aerogel glasses of 1\% and 2\% of solid density.  NMR occurs
over a range of frequency extending from the Larmor frequency to
higher values, but strongly peaked at the Larmor value.  This
behaviour together with the magnetic field independence of the
effective superfluid density provides convincing evidence for a
B-phase state with an ${\bf {\hat n}}$ texture, in our spherical
geometry, governed by the same energetic considerations as for bulk
superfluid $^3$He-B.

\end{abstract}
\narrowtext

Since the discovery\cite{Porto95,Halperin95} of superfluidity of
$^3$He contained within aerogel, this system has been
regarded\cite{Thuneberg98} as providing an excellent model for
investigating the effect of impurity scattering on Fermi superfluids
with non s-wave Cooper pairing.  One of the main theoretical
expectations\cite{Thuneberg98} was that the introduction of scattering
would increase the stability of the isotropic B-phase relative to that
of the strongly anisotropic A-phase.  It was surprising therefore that
previously published NMR experiments on the superfluid
phase\cite{Halperin95,Barker98} were characteristic of an equally spin
paired (ESP) state, of which the A-phase is the prime example, even at
pressures below those at which the A-phase is stable in bulk.  Our
experiments were performed at much smaller magnetic fields than those
used in these previous experiments and they provide convincing
evidence for the occurrence of B-phase under these circumstances.  We
note that our aerogel samples came from the same source as those
used in the measurements in reference \cite{Halperin95} and, since the
observed superfluid transition temperatures for our 2\% aerogel were
very similar to those observed in reference \cite{Halperin95}, there
is no reason to suspect that the difference in NMR behaviour results
from a significant difference in aerogel microstructure.

Some details of our experiment along with measured
values of the transition temperature and superfluid density have
already been published\cite{Alles98} so we do not repeat these here.
We studied two samples of aerogel, of 1\% and 2\% of solid
density, each contained within a glass sphere of inner diameter 8\,mm.
The sphere was supported on a Be/Cu capillary which acted as the fill
line and also provided the torsion constant for the measurements of
superfluid density.  A pair of coils each with 20 turns of filamentary
superconducting wire were driven at fixed frequency and current to
produce an rf field transverse to the
dc field and the NMR signal was detected by another
pair of transverse coils wound from the same wire each with 100 turns,
arranged as orthogonally as possible to the `drive' coils.  
The `detection' coils were tuned to a frequency of
approximately $165\,{\rm kHz}$ by a $2200\,{\rm pF}$ polystyrene
capacitor and the signal was observed with the help of a $4.2\,{\rm
K}$ GaAs preamplifier\cite{Ruutu94}  followed by a further
differential preamplifier and lockin detector; the quality factors of
the tuned circuit were of order 3000 and 5000 for the measurements on
the 2\% and 1\% samples respectively.  We observed cw NMR
by sweeping the dc field at constant rf drive frequency. 

\begin{figure}[htb]
\includegraphics[angle=0,width=75 mm]{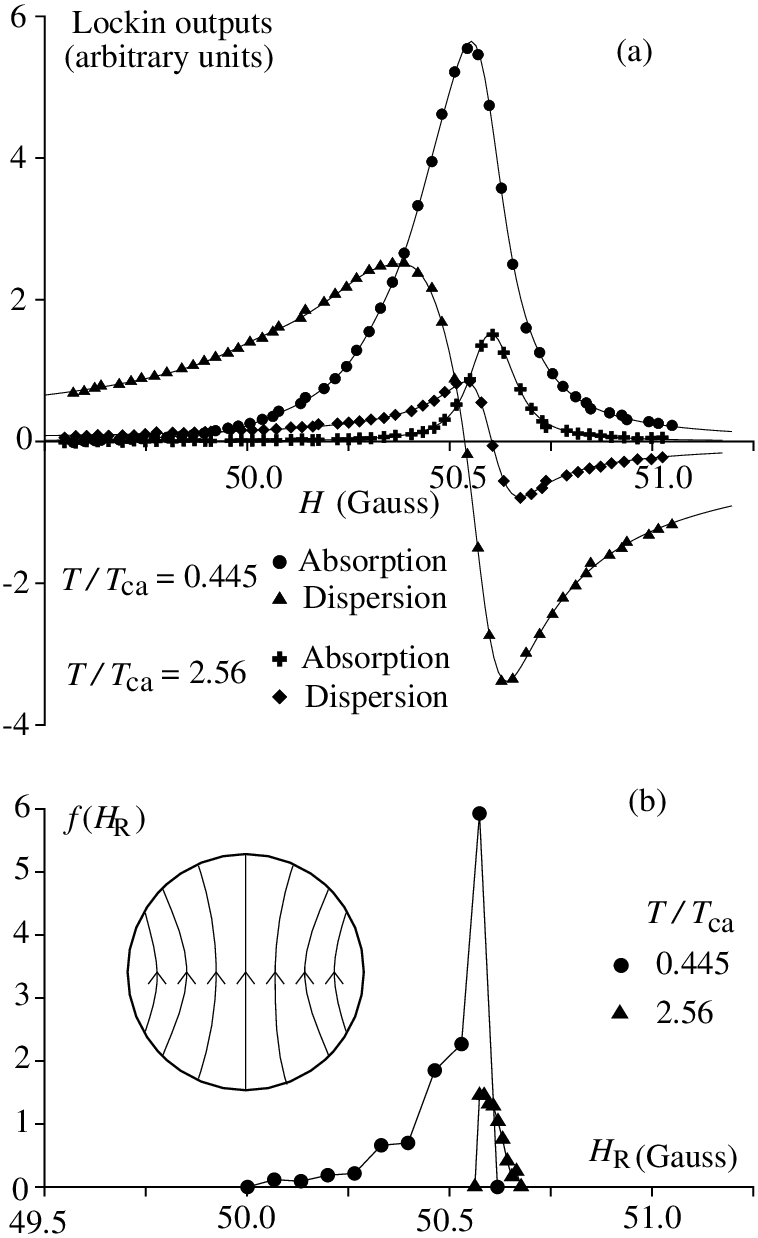}
\caption{(a) Absorption and dispersion NMR signals from 2\% aerogel at
14.9\,bar.  (b) The spectrum functions $f(H_R)$ which give the fits
indicated by the solid lines on (a). The inset on (b) shows, schematically,
the projection into the $r-z$ plane of the flare-out ${\bf{\hat n}}$
texture believed to be responsible for the spectra.}
\label{Fig1}
\end{figure}

We show in Fig.\,\ref{Fig1}(a) the observed NMR absorption and
dispersion for field sweeps through the resonance for 2\% aerogel at
14.9\,bar pressure; sweeps for temperatures above and below the
superfluid transition temperature $T_{\rm ca}$ are shown.
The large
increase in signal strength at the lower temperature is associated
with the contribution from solid $^3$He atoms on the surfaces of the
silica strands.  As in reference \cite{Halperin95} we assume that
rapid exchange of $^3$He atoms
between liquid and solid results in a single NMR signal with a strength
$I$ (integrated absorption) proportional to the sum of the
liquid and solid magnetizations, which can be fitted to
\begin{equation}
I=A+{B\over{T+T_0}},
\label{one}
\end{equation}
where $A$ represents a temperature-independent contribution from the
liquid; for the normal state, the measured variation of $A$ with
pressure is the same as that of the susceptibility of bulk liquid.
The Curie-Weiss term is the contribution from the solid.  $T_0$ is of
order $-0.4\,{\rm mK}$ and almost pressure-independent.  The values
obtained for $B$ correspond to an areal density of order
$10^{19}\,{\rm m}^{-2}$ of solid $^3$He atoms, i.e.  about one
monolayer; $B$ increases with pressure by $\sim30\%$ between 0 and
29\,bar.  We discuss below the extent to which Eq.\,(\ref{one}) is
valid for $T<T_{\rm ca}$.

From Fig.\,\ref{Fig1}(a) we see that NMR in the superfluid state
occurs over a range of field negatively shifted with respect to the
normal state Larmor value.  To analyse this spectrum we assume that it
arises from a distribution of local Lorentzian oscillators within the
aerogel with $f(H_R)$ giving the contribution to the spectrum from
oscillators of resonant field $H_R$; we thus write the NMR signal as
\begin{equation}
{\rm Signal}=\int\frac{f(H_R){\rm d}H_R}{H_R-H+iH_B/2},
\label{two}
\end{equation}
where we assume that the linewidth $H_B$ is constant across the
spectrum.  In order to fit the data we assume that the distribution
function $f(H_R)$ is zero for $H_R\leq H_1$ and for $H_R\geq H_2$, and
that, for $H_1<H_R<H_2$, it can be represented by its value at nine
values of $H_R$; these were normally equally spaced but when the
spectrum was very broad it proved desirable to limit the spacing of
the two intervals nearest the Larmor field to be no greater than
$50\,{\rm mGauss}$ in order to obtain a satisfactory fit to this end
of the spectrum.  Using nine points to fit the spectrum allows us to
obtain a reasonably detailed shape for $f(H_R)$ without introducing
spurious detail associated with the noise on the data.
Fig.\,\ref{Fig1}(b) shows the `nine-point' spectra used to obtain the
fits shown by solid lines on Fig.\,\ref{Fig1}(a); the values of $H_1$
and $H_2$ were fitting parameters. The finite width of the spectrum
for the normal state
can be associated with field inhomogeneity; for the superfluid state
the spectrum confirms that only negative field (i.e.  positive
frequency) shifts are observed and that the spectrum is strongly
peaked towards the Larmor field.  These features of our data are
common to all pressures where superfluid transitions were observed;
i.e. from 1.5\,bar to 29.3\,bar for the 1\% aerogel and from
4.8\,bar to 29.3\,bar for the 2\% aerogel.  The linewidths $H_B$
obtained from our fits show a slight increase with decreasing
temperature which is not affected by the superfluid transition.

The most likely explanation for NMR to occur over a spectrum of
frequencies is a superfluid phase with a space-dependent texture.  We
now present three pieces of evidence that our observations
correspond to superfluid B-phase with a space dependent ${\bf {\hat n}}$
texture rather than A-phase with spatially dependent $\bf{\hat l}$ and
$\bf{\hat d}$ textures.  First this would explain why only positive
frequency shifts are observed since the transverse NMR frequency
$\omega$ for liquid $^3$He-B with the ${\bf {\hat n}}$ vector at
an angle $\beta$ to the applied field, as given
by\cite{Vollhardt90}
\begin{equation}
2\omega^2=\left(\omega_L^2+\Omega_B^2\right)
+\sqrt{\left(\omega_L^2+\Omega_B^2\right)^2
-4\omega_L^2\Omega_B^2\cos^2\beta},
\label{three}
\end{equation}
is always greater than the Larmor frequency $\omega_L$; here
$\Omega_B$ is the B-phase longitudinal frequency.  Spatially varying
textures in the A-phase are likely to lead to negative as well as
positive shifts\cite{Barker98}.  Second, the shape of the NMR
spectrum is qualitatively that expected for the $\hat{\bf n}$ texture
in a spherical geometry.  In particular the strong peak of the
spectrum at the Larmor frequency arises because the favoured alignment
of $\hat{\bf n}$ in a field is $\beta=0$ which according to
Eq.\,(\ref{three}) gives an unshifted resonant frequency,
$\omega=\omega_L$; it is difficult to conceive of an A-phase texture
which gives a peak at the Larmor frequency but which has only
positive shifts.  Thirdly, one would expect a spatially varying
$\bf{\hat l}$ texture in the A-phase to depend on magnetic field
and thus lead to variations in effective superfluid density with field
through the anisotropy of the superfluid density.  In our experiments
we regularly swept the magnetic field from positive to negative
values\cite{explanation2} but the associated small change in resonant
frequency ($\sim4\,{\rm mHz}$) of our torsional oscillator was
unaffected by the superfluid transition, thus ruling out changes in
the effective value of $\rho_{\rm s}/\rho$ greater than about 0.0003.
The possibility that the field independence of the effective
superfluid density can be associated with a random dipole-unlocked
orbital texture pinned by the aerogel is inconsistent with the
observed NMR spectrum.

We now demonstrate that we can explain the observed spectra
quantitatively if 
we assume that the ${\bf {\hat n}}$ texture in aerogel is determined
by the same considerations as for bulk $^3$He-B.  As the
characteristic lengths for
B-phase textures (see below) are large,  we can ignore spatial
variation of $\bf{\hat n}$ on the scale of the aerogel microstructure
and determine the texture by minimizing the energy\cite{Brinkman78}
\begin{eqnarray}
E=\int\{-b[({\bf {\hat s}}\cdot{\bf {\hat n}})^2
+\nu({\bf {\hat s}}\cdot{\bf {\hat n}})^4]-
d({\bf {\hat s}}\cdot{\bf R}\cdot{\bf H})^2\}{\rm d}S\nonumber\\
+\int\left(f_{\rm bend}
-a\left({\bf {\hat n}}\cdot{\bf H}\right)^2\right){\rm d}V
\label{four}
\end{eqnarray}
where $a$, $b$, $c$, $d$ and $\nu$ (we take
$\nu=-5/18$\cite{Vollhardt90}) are temperature dependent coefficients,
${\bf{\hat s}}$ is the surface normal to the sphere, the matrix
${\bf R}$ represents a rotation about ${\bf{\hat n}}$ by
$\cos^{-1}(-1/4)\approx104^\circ$ and the integrals are over the
surface and volume of the sphere respectively.  For the bending
energy density $f_{\rm bend}$, we take the Ginzburg-Landau form
\begin{eqnarray}
f_{\rm bend} & = & c\{16(\nabla\times{\bf{\hat n}})^2
-5[{\bf{\hat n}}\cdot(\nabla\times{\bf{\hat n}})]^2
+13(\nabla\cdot{\bf{\hat n}})^2
\nonumber\\ & - & 2\sqrt{15}\nabla\cdot{\bf{\hat n}}
[{\bf{\hat n}}\cdot(\nabla\times{\bf{\hat n}})]\nonumber\\
 & + & 16\nabla\cdot[({\bf{\hat n}}\cdot\nabla){\bf{\hat n}}
-{\bf{\hat n}}(\nabla\cdot{\bf{\hat n}})]\}/13.
\label{five}
\end{eqnarray}

We assume that the ${\bf{\hat n}}$ texture has rotational symmetry
about the field direction (the $z$ axis) with cylindrical polar
components $n_r=\sin\beta\cos\alpha$, $n_\phi=\sin\beta\sin\alpha$,
$n_z=\cos\beta$.  To make the texture calculation tractable we use
the variational forms
\begin{eqnarray}
\tan\beta=\frac{2(z-p_1)r}{z(z-2p_1)-r^2+p_2^2},\label{sixa}\\
\tan\alpha=p_3+p_4z+p_5z^2+p_6r^2,
\label{sixb}
\end{eqnarray}
for $\beta$ and $\alpha$; the variational parameters $p_1$-$p_6$ are
chosen to minimise $E$ for given values of the coefficients in the
energy of Eq.\,(\ref{four}).  For $p_1=p_4=0$ the texture of
Eqs.\,(\ref{sixa}) and (\ref{sixb}) corresponds to the texture shown
in the insert on
Fig.\,\ref{Fig1}(b), with ${\bf{\hat n}}\parallel{\bf H}$
(i.e. $\beta=0$) on the equatorial plane and on the line joining the
north and south poles; the second surface term in Eq.\,(\ref{four})
causes ${\bf{\hat n}}$ to point into and out of the paper in the
south-east and north-east quadrants respectively.  Our investigations of
possible ${\bf{\hat n}}$ textures in a sphere suggest that this
variation of the `flare-out' texture in a cylinder\cite{Spencer82} has
the lowest energy for a spherical container.  Finite values of $p_1$
and $p_4$ correspond to moving the $\beta=0$ plane away from the
equator but this always caused an increase in $E$ in practice.

To relate the ${\bf{\hat n}}$ texture to the observed spectrum, we
solved the equations which describe NMR for the
situation of rapid exchange between solid and B-phase liquid. The
calculation, details of which will be published elsewhere, shows
that, at a constant measuring frequency $\omega$, the
resonance occurs at a field $H$ given by
\begin{equation}
(\gamma H)^2=\omega_L^2=\omega^2
\left(\frac{\omega^2-\mu\Omega_B^2}{\omega^2-
\mu\Omega_B^2\cos^2\beta}\right),
\label{seven}
\end{equation}
where $\mu=M_{\rm liqu}/(M_{\rm liqu}+M_{\rm sol})$.  $M_{\rm liqu}$
and $M_{\rm sol}$ are the magnetizations of the liquid and solid
respectively.  In the limit $\mu=1$ Eq.\,(\ref{seven}) reduces to
Eq.\,(\ref{three}) and in the limit $\mu\Omega_B^2\ll\omega^2$ it
reduces to $\omega\approx(1-\mu)\omega_{\rm sol}+\mu\omega_{\rm liqu}$
as used in interpreting previous NMR measurements\cite{Halperin95} on
superfluid $^3$He in aerogel. 

\begin{figure}[htb]
\includegraphics[angle=0,width=80 mm]{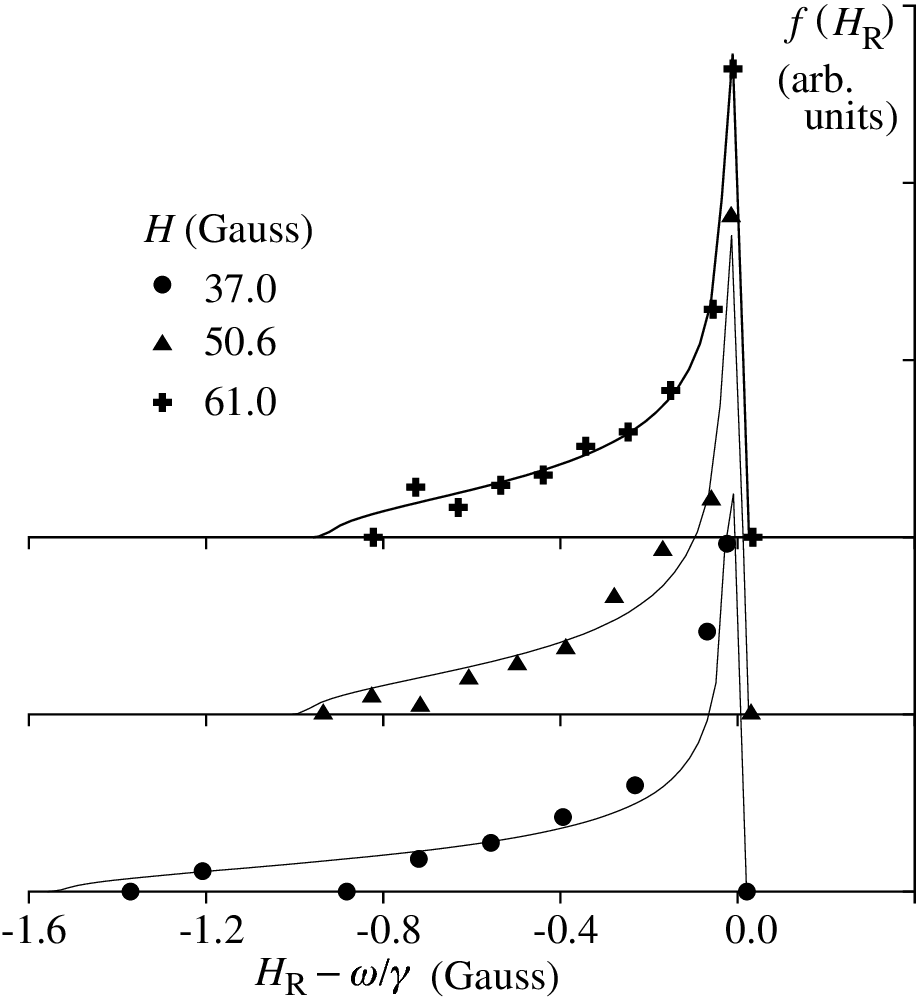}
\caption{Experimental (circles, triangles and crosses) and fitted NMR
spectra (continuous lines) for 1\% aerogel at 11.6\,bar and $T/T_{\rm
ca}=0.63$.  There was some evidence of structure in some of the
observed spectra, which may indicate the existence of textural spin
waves\protect\cite{Osheroff77}.}
\label{Fig2}
\end{figure}

To fit our experimental spectra we note that the ${\bf{\hat n}}$
texture depends only the ratios $a/c$, $b/c$ and $d/c$ and that the
only other unknown is $\Omega_B$.  We take $a/c=(q_1+q_2t)/t$,
$b/c=(q_3+q_4t)/t^{1/2}$, $d/c=(q_5+q_6t)/t^{1/2}$ and
$\Omega_B^2=(q_7t+q_8t^2)$, where $t=1-T/T_{\rm ca}$; these have the
expected temperature dependences for $T\rightarrow T_{\rm ca}$ and
allow adequately for departures from these asymptotic dependences at
lower temperatures.  Adjustment of $q_1$-$q_8$ enables us to fit all
our spectra at a single pressure within experimental error.  To
illustrate this we show in Fig.\,\ref{Fig2} typical examples of fitted
spectra obtained for 1\% aerogel at 11.6\,bar.
The data at this
pressure provided a stringent test for our explanation since
measurements were made at three different magnetic fields (37.0, 50.6
and 61.0 Gauss)\cite{explanation3}.

\begin{figure}[htb]
\includegraphics[angle=0,width=80 mm]{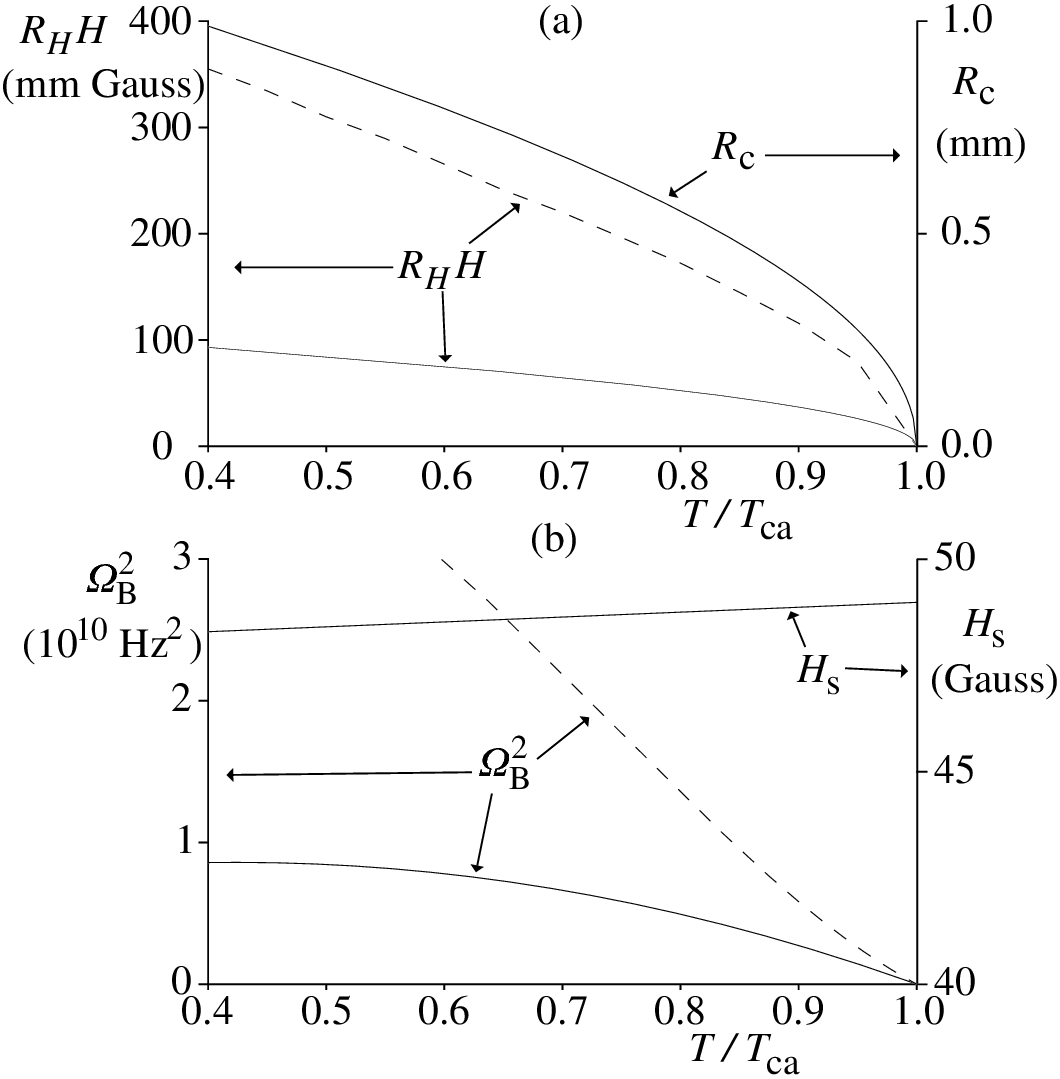}
\caption{The values of (a) $R_HH$ and $R_{\rm c}$, and (b) 
$\Omega_B^2$ and $H_{\rm s}$, used to obtain fits to NMR spectra for
1\% aerogel at 11.6\,bar are shown by continuous lines. The dashed
lines are for bulk $^3$He-B at 10.2\,bar.} 
\label{Fig3}
\end{figure}

The resulting values of the free energy coefficients are best
presented in terms of two characteristic distances, $R_H$ and $R_c$,
and one characteristic field $H_s$ defined by:  $R_HH=(c/a)^{1/2}$;
$R_c=c/b$; $H_s=(b/d)^{1/2}$\cite{Brinkman78}.  The significance of
these quantities in determining the relative importance of
different terms in the energy can be seen from Eqs.\,(\ref{four}) and
(\ref{five}).  The values obtained for $R_HH$, $R_c$, $H_s$ and
$\Omega_B^2$ are shown in Fig.\,\ref{Fig3} along with values of $R_HH$
and $\Omega_B^2$ for bulk $^3$He-B at 10.2\,bar\cite{Hakonen89}.
As can be seen from Eq.\,(\ref{seven}), only the product $\mu\Omega_B^2$
can be determined from NMR spectra; to obtain $\Omega_B^2$
we used values of $\mu$ deduced from fits to Eq.\,(\ref{one}) over the
whole temperature range, a procedure which assumes that the liquid
susceptibility is constant for $T<T_{\rm ca}$.  At 11.6\,bar we could
not detect departures from Eq.\,(\ref{one}) associated with reduction
of the B-phase susceptibility below that of normal $^3$He.  The
departures are expected to be small because of the strong supression
of the energy gap in aerogel and, as indicated in
Ref.\,\cite{Barker98}, it is difficult to evaluate the integrated
absorption for a broad NMR spectrum with sufficient accuracy to
identify small changes in the liquid contribution.  There was evidence
for departures from Eq.\,(\ref{one}) at the highest pressures studied
where the effect is larger and the solid contribution relatively less
important.  A reduction in liquid susceptibility corresponds to a
smaller value of $\mu$ and hence to a slightly larger value of
$\Omega_B^2$ than that shown on Fig.\,\ref{Fig3}(b).  Although this
effect would have to be included in any quantitative comparison with
theoretical calculations, the qualitative conclusion from
Fig.\,\ref{Fig3}(b) remains valid that the depression of the energy
gap of $^3$He in aerogel causes the value of $\Omega_B^2$ to be
substantially less than that for bulk $^3$He-B.  Our values of $R_HH$
are also smaller than bulk value as are our values of
$d/a$\cite{Korhonen90}.  The latter are not shown on
Fig.\,\ref{Fig3} since the ratio $a/c$ used to deduce
them is the least well determined of the fitting parameters and it was
desirable that it should be used only once (in calculating
$R_HH$) in presenting a complete set of fitting parameters; the
poor determination of $a$ may arise because its role in aligning
$\bf{\hat n}$ parallel to $H$ is partly usurped by the choice of
the variational form (\ref{sixa}) for $\beta$.  `Bulk' values of
$R_{\rm c}$ and $H_{\rm s}$ are not shown in Fig.\,\ref{Fig3} since we
are not aware of previous measurements of $b$ for bulk $^3$He at a
relevant pressure.

Explanation of the values of $R_HH$, $R_c$, $H_s$ and $\Omega_B^2$ and
their dependence on pressure and aerogel density provides a problem
for theoreticians.  We believe that we have presented convincing
evidence for a B-phase superfluid state of $^3$He in aerogel.  It is
likely that the ESP behaviour observed in Refs.\,\cite{Halperin95} and
\cite{Barker98} is associated with the substantially higher field used
in these experiments; further evidence that this is the case is
provided by the observation that replacement of the very magnetic
solid $^3$He layer by $^4$He produced more B-phase-like behaviour in
both these experiments\cite{Barker98,Sprague96}.

This work was supported by EPSRC through Research Grants GR/K58234,
GR/K59835 and GR/M15750.  We are grateful to Moses Chan and Norbert
Mulders for providing the aerogel samples used in this work and to
Jaakko Koivuniemi for providing the 4.2\,K preamplifier.  We
acknowledge useful communications with Brian Cowan, Andrei Golov,
Henry Hall, Bill Halperin, Doug Osheroff, Jeevak Parpia, Bob
Richardson, Don Sprague and Erkki Thuneberg.


\begin{references}
\bibitem{explanation2}We made NMR measurements in
both positive and negative fields in order to correct for the small
changes in dc magnetic field associated with flux creep in our
demagnetization solenoid.
\bibitem{explanation3}The NMR signal was much smaller away
from the resonant frequency of our detection circuit, but measurements
were possible.
\end{references}
\end{document}